\documentclass[prd,aps,12pt,preprintnumbers,amssymb,amsmath,amsfonts,nofootinbib]{revtex4-2}

\usepackage{latexsym}
\usepackage{ulem}
\usepackage[utf8]{inputenc}
\usepackage{cancel}

\newcommand{\ba}{\begin{eqnarray}}
\newcommand{\ea}{\end{eqnarray}}

\newcommand{\be}{\begin{equation}}
\newcommand{\ee}{\end{equation}}

\usepackage[x11names,dvipsnames,hyperref]{xcolor}

\usepackage{slashed}
\usepackage{pifont}
\usepackage{dcolumn}% Align table columns on decimal point
\usepackage{graphics}                                             \usepackage{braket}                             
\usepackage[dvips]{graphicx}
\usepackage{multirow}
\usepackage{makecell}  
\usepackage{subcaption}

\begin{document}
                                                                                
\date{\today}
\title{Femtoscopy of $D$ mesons and light mesons \\ upon unitarized effective field theories
}

\author{Juan M. Torres-Rincon$^1$, \`Angels Ramos$^1$ and Laura Tolos$^{2,3,4}$ }

 \affiliation{$^1$Departament de F\'isica Qu\`antica i Astrof\'isica and Institut de Ci\`encies del Cosmos (ICCUB), Facultat de F\'isica,  Universitat de Barcelona, Mart\'i i Franqu\`es 1, 08028 Barcelona, Spain}
  \affiliation{$^2$Institute of Space Sciences (ICE, CSIC), Campus UAB, Carrer de Can Magrans, 08193, Barcelona, Spain}
 \affiliation{$^3$Institut d'Estudis Espacials de Catalunya (IEEC), 08034 Barcelona, Spain}
 \affiliation{$^4$Frankfurt Institute for Advanced Studies, Ruth-Moufang-Str. 1, 60438 Frankfurt am Main, Germany}
% \keywords{}
\date{\today}

\begin{abstract}
Hadron femtoscopy has turned into a powerful tool for accessing space-time information of heavy-ion collisions as well as for studying final-state interactions of hadrons. Recently, heavy-flavor femtoscopy has become feasible using the ALICE detector at the LHC. We compute the correlation function of $D$ mesons and light mesons using an off-shell $T$-matrix approach to access the two-meson wave function, and predict the correlation functions involving charged $D^+, D^{*+},D_s^+$ and $D_s^{*+}$ with $\pi^\pm$ and $K^\pm$. From the obtained results---all of them accessible in $p+p$ collision experiments---we point up the case of $D^+ \pi^-$, which is sensitive to the lower state of the two-pole $D_0^* (2300)$ system. The presence of such poles imprints a depletion on the correlation function, which could potentially be detected in experiments. While preliminary ALICE data do not show evidence of this effect, we suggest to look into the $D_s^+ K^-$ system to explore the higher pole of the $D_0^* (2300)$, as the depletion in the correlation function is more pronounced. Using heavy-quark spin symmetry we also propose exploring the effect of the two poles of the $D_1(2430)$ and predict similar structures in the correlation functions of the $D^{*+} \pi^-$ and $D_s^{*+} K^-$ pairs.
\end{abstract}

\maketitle

%\tableofcontents

\section{Introduction}

Understanding the strong interaction among hadrons is still an open question in nuclear physics. Quantum Chromodynamics (QCD) is the basic theory of the strong interaction. While QCD is well tested at distances much shorter than the size of the nucleon and many processes at high energies can be described at the quark level by means of perturbative QCD, this perturbative approach fails when the distance between quarks is comparable to the nucleon size. In this low-energy regime, QCD becomes a strongly coupled theory and the low-energy processes between hadrons are not yet well described theoretically and are often difficult to access experimentally.

In the past the interaction between hadrons has been extracted experimentally using scattering experiments at low energies (below the nucleon mass) with both stable and unstable beams. There is a large amount of scattering data for nucleon-nucleon reactions \cite{Arndt:2007qn,NavarroPerez:2013usk}, but as we access heavier degrees of freedom, such as strangeness or charm, the experimental realization is more challenging. The extremely short lifetime of hadrons containing heavier quarks than up or down makes it very difficult to perform scattering experiments and the knowledge of the interaction among heavy hadrons is obtained from reactions where hadrons are produced in the final state. 

In this context, in the last decades femtoscopy has emerged as a interesting tool to study reactions among hadrons \cite{Fabbietti:2020bfg}. Femtoscopy techniques consist in measuring the hadron-hadron correlation in momentum space, which can be obtained as the ratio of the distribution of relative momenta for pairs produced in the same collision and in different collisions (mixed events).  The development of this technique to study the hadron-hadron interaction was pioneered in \cite{HADES:2016dyd,Shapoval:2014yha} and further developed by the STAR Collaboration for $\Lambda \Lambda$ \cite{STAR:2014dcy}, $\bar p \bar p$ \cite{STAR:2015kha} and $p \Omega^-$ \cite{STAR:2018uho} correlations. In the recent years, the ALICE Collaboration has analyzed correlations for different systems, such as for kaon-kaon \cite{ALICE:2017jto,ALICE:2017iga,ALICE:2018nnl,ALICE:2021ovd}, $pp$ \cite{ALICE:2018ysd}, $\Lambda K$  \cite{ALICE:2020wvi}, $pK^+$ and $pK^-$ \cite{ALICE:2019gcn}, $p \Lambda$ \cite{ALICE:2018ysd,ALICE:2021njx}, $p \Sigma ^0$ \cite{ALICE:2019buq}, $\Lambda \Lambda$ \cite{ALICE:2018ysd,ALICE:2019eol}, $p \Xi^-$ \cite{ALICE:2019hdt}, $p \Omega^-$ \cite{ALICE:2020mfd}, $K^{\pm} \pi^{\pm}$ \cite{ALICE:2020mkb}, $p \phi$ \cite{ALICE:2021cpv}, as well as $p \bar p$, $p \bar \Lambda$, $\bar p \Lambda$ and $\Lambda \bar\Lambda$ \cite{ALICE:2019igo,ALICE:2021cyj}. 

More recently, femtoscopic studies have moved to the charm sector as the ALICE Collaboration has measured $p D^-$ and $\bar p D^+$ correlations in high-multiplicity $pp$ collisions at 13 TeV \cite{ALICE:2022enj}. Also, recent results are being reported by the ALICE Collaboration for $D^{(\pm)} \pi^{(\pm)}$ and $D^{(\pm)} K^{(\pm)}$ reactions \cite{Grosa2022,Fabbietti2022,Battistini2023,ALICE:prelim}, thus giving new insights into the hadron-hadron interactions with charm content.

In view of the present developments in femtoscopy in the charm sector and the expected future advances, some theoretical analyses of the hadron-hadron correlation function including charmed hadrons have been performed. The femtoscopic correlation functions for the $D^0 \bar D^{*0}$ and $D^+ D^{*-}$ channels have been obtained in \cite{Kamiya:2022thy} so as to analyze the nature of the $X(3872)$, as well as the ones for $D^0 D^{*+}$ and $D^+ D^{*0}$ to address the features of the newly discovered $T_{cc}$ state \cite{Kamiya:2022thy,Vidana:2023olz}. Also, the authors in \cite{Liu:2023wfo} propose to determine the spins of $P_c(4440)$ and $P_c(4457)$ states by measuring the $\Sigma_c^+ \bar D^{(0)*}$ correlation functions.  Moreover, there are calculations in the isospin $I=1/2$ and strange $S=0$ sector for the $D \pi$, $D \eta$ and $D_s \bar K$ correlation functions to determine the two-pole structure of the $D_0(2300)$ \cite{Albaladejo:2023pzq}, as well as theoretical studies in the $I=0$ and $S=1$ sector for the $D K$ \cite{Albaladejo:2023pzq,Liu:2023uly,Ikeno:2023ojl} and $D_s \eta$ \cite{Ikeno:2023ojl} correlation functions,  where the $D_{s0}^* (2317)$ has a prominent role \cite{Albaladejo:2023pzq,Liu:2023uly,Ikeno:2023ojl}. 

In the present work we follow the previous analyses in the $I=1/2,S=0$ and $I=0,S=1$ sectors of Refs.~\cite{Albaladejo:2023pzq,Liu:2023uly}. However, these previous theoretical works only consider meson-meson correlations with one charged meson, whereas experimentally the correlation functions have been obtained for a pair of charged mesons. Therefore, in this work we study the correlation functions for $D^{+} \pi^{+}$ and $D^{+} \pi^{-}$  (and associated charge conjugates) together with the ones for $D^{+} K^{+}$ and $D^{+} K^{-}$ (and associated charge conjugates) with the aim of comparing our results with the ALICE experimental outcome~\cite{Grosa2022,Fabbietti2022,Battistini2023,ALICE:prelim}. We moreover predict novel correlation functions for channels involving $D_s$, $D^*$ and $D_s^*$ mesons. For that purpose, our study employs heavy-light meson-meson unitarized effective interactions derived from an off-shell $T$-matrix calculation in a coupled-channel basis. In that framework, the two-pole $D_0^*(2300)$ and the $D_{s0}^* (2317)$ states are dynamically generated by the heavy-light meson-meson scattering. Concerning the $D^*$ meson, analogous states in the $J=1$ sector also appear, namely the two-pole $D_1(2430)$ and the bound state $D_{s1}(2460)$. 

The correlation functions of heavy-light mesons are then calculated, accounting for the Coulomb interaction in the relevant channels. The final goal of our study is to  describe the heavy-light meson-meson interactions in the different charm sectors, while determining the role of the two-pole $D_0^*(2300)$ and $D_{1}(2430)$ in the correlation functions.

The paper is organized as follows. In Sec.~\ref{sec:femtoscopy} we review the basic formalism to compute hadron-hadron correlation functions using the off-shell $T$-matrix approach to access the pair wave function. We provide some details on the strong and Coulomb forces relevant to describe the charged hadron-hadron interactions.
In Sec.~\ref{sec:results} we show our results on the correlation functions for charged channels involving $D$ mesons (Sec.~\ref{sec:D}), as well as $D_s$ (Sec.~\ref{sec:Ds}) and $D^*$ and $D_s^*$ mesons (Sec.~\ref{sec:Dstar}). In Sec.~\ref{sec:conclu} we present our conclusions and outlook.

\section{Femtoscopy formalism~\label{sec:femtoscopy}}

In this section we present the formalism employed to obtain the correlation function of a pair of charged mesons. We consider channels composed by a charmed meson (denoted as $D$ meson) and a light pseudoscalar, focusing on the cases recently measured by the ALICE collaboration~\cite{Grosa2022,Fabbietti2022,Battistini2023,ALICE:prelim}, namely $D^+\pi^\pm, D^+ K^\pm$ (and their charge conjugated pairs $D^-\pi^\mp, D^- K^\mp$) and giving predictions for the cases $D_s^+ \pi^\pm, D_s^+ K^\pm$ ($D_s^- \pi^\mp, D_s^- K^\mp$) involving the charged $D_s$ mesons, as well as all their vector companions ($D^*$ and $D_s^*$ mesons). The pseudoscalar channels are listed in boldface in Table~\ref{tab:chann}, together with their coupled partners in each of the strangeness $S$ and charge $Q$ sectors.

\subsection{Pair correlation function}

The femtoscopic correlation function of a particular hadron pair with relative momentum $\boldsymbol{q}$ is obtained from the Koonin-Pratt formula~\cite{Koonin:1977fh,Pratt:1990zq},

\begin{equation}
\label{eq:KP}
C(\boldsymbol{q})=\int d^3r  \sum_i w_i \ S_i(\boldsymbol{r}) \ | \Psi_i(\boldsymbol{q};\boldsymbol{r}) |^2 \ ,
\end{equation}
where $\Psi_i(\boldsymbol{q};\boldsymbol{r})$ is the wave function converting the pair in the $i{\rm th}$ channel into the asymptotically measured one ($i\rightarrow f$)\footnote{All wave functions are associated to the asymptotic final state $f$. Their subindexes exclusively denote the initial channels.}, $w_i$ stands for the weight representing the strength in which the pair $i$ is created in the collision and $S_i(\boldsymbol{r})$ is the normalized source function representing the distribution of relative distances $\boldsymbol{r}$ at which particles are emitted. 

The asymptotic channels considered in the present work are composed by a pair of charged mesons that feel the effect of both the strong and the Coulomb forces. Considering only the $s$-wave component of the strong interaction, which is a valid approximation at low momenta, the wave function for the asymptotic pair of charged mesons (denoted with the subscript $f$) is given by 
\begin{equation}
\Psi_f(\boldsymbol{q};\boldsymbol{r})= \Phi_f^{\rm C}(\boldsymbol{q};\boldsymbol{r}) - \Phi^{\rm C}_{0\, f}(q\,r) + \varphi_f(q;r) \ ,  \label{eq:decomp}
\end{equation}
where $q\equiv |\boldsymbol{q}| $ and $r\equiv |\boldsymbol{r}|$, while the transition wave function of the other channels ($j\ne f$) reads
\begin{equation}
%\Psi_j(\boldsymbol{q};\boldsymbol{r})=  \varphi_j(q;r) \ ,
\Psi_j(\boldsymbol{q};\boldsymbol{r})=  \varphi_j(q;r) \ ,
\label{eq:phi_j}
\end{equation}
where $\varphi(q;r)$ in these equations stands for the $s$-wave wave function including the effect of the strong interaction and, for channels composed by a pair of charged mesons, also that of the Coulomb force. The function  $\Phi^{\rm C} $ stands for the complete Coulomb wave function and $\Phi^{\rm C}_{0}$ by its $s$-wave component~\cite{joachain1975quantum}. 

Guided by experimental evidence, the source function can be taken spherically symmetric and independent of the initial channel $i$ evolving into the asymptotic one $f$. Under these conditions, the correlation function becomes
\begin{equation}
\label{eq:corr}
C (q)=\int d^3r  S(r) \ |\Phi^{\rm C}_f(\boldsymbol{q};\boldsymbol{r}) |^2 
+\int 4\pi r^2\, dr  S(r) \left[ \sum_i w_i \ | \varphi_i(q;r) |^2 - | \Phi^{\rm C}_{0\, f}(q\,r ) |^2 \right] \ .
\end{equation}
In our calculations we employ a Gaussian source function,
\begin{equation}
S(r)=\frac{1}{(2\sqrt{\pi}R)^3}\,{\rm exp} \left( -\frac{r^2}{4R^2} \right) ,
\end{equation}
with $R=1$ fm being the typical size of the primary source for the systems considered here~\cite{Battistini2023}. The source is normalized as
\begin{equation}
\int d^3r \ S(r)=1 \ .
\end{equation}
Finally, given the lack of experimental information, we set all the weights $w_i$ of Eq.~(\ref{eq:KP}) equal to unity.

\subsection{Wave function and $T$ matrix}

In this work we are going to obtain the wave function from the $T$-matrix amplitude, according to the Lippmann-Schwinger equation,
\begin{equation}
    \ket{\Psi} = \ket{\Phi}  + \frac{1}{E-\hat{H}_0+{\rm i}\eta} T \ket{\Phi} \ , 
\end{equation}
where $\ket{\Phi}$ is the free wave function, while $\ket{\Psi}$ contains the effect of the interaction. From the former equation, and assuming an interaction projected in $s$-wave, one obtains~\cite{joachain1975quantum}, 
\begin{equation}
    \varphi_{i}(q; r) = j_0(q r) \delta_{i f} +
    \int_0^\infty \frac{4\pi q^{\prime\,2}\,d q^\prime}{(2\pi)^3} \frac{T_{i f}(q^{\prime},q;\sqrt{s}) \, j_0(q^{\prime} r)}{2 \omega_{H, i}\, 2 \omega_{\phi, i} (\sqrt{s}- \omega_{H, i}-\omega_{\phi, i} + {\rm i}\eta )} \ ,
\end{equation}
where $j_0(qr)$ is the spherical Bessel function, and $T_{if}$ is the scattering amplitude that is derived from a Bethe-Salpeter equation.
\begin{equation}
    T_{if} (q',q; \sqrt{s}) = V_{if}(q',q;\sqrt{s}) + \sum_l \int_0^\infty \frac{4\pi k^2 dk}{(2\pi)^3} \frac{V_{il}(q',k;\sqrt{s}) \,T_{lf} (k,q;\sqrt{s})}{2 \omega_{H,l} \, 2 \omega_{\phi,l} \ (\sqrt{s} - \omega_{H,l}-\omega_{\phi,l} + {\rm i}\eta ) }  \ ,
    \label{eq:Tmat}
\end{equation}
with $V_{if}$ being the $s$-wave projected interaction kernel containing the effect of the strong force and, for diagonal transitions involving a pair of charged mesons, also that of the Coulomb interaction. In the above equations $\omega=\displaystyle\sqrt{m^2+p^2}$ is the relativistic energy of a meson of mass $m$ and momentum $p$.

Equation~(\ref{eq:Tmat})---also denoted as the half off-shell $T$-matrix equation---will be solved numerically for each $(S,Q)$ sector. The typical ultraviolet divergence in the integrals is cured by inserting a Gaussian form factor,
\begin{equation}
f(q,q')=\exp \left( - \frac{q^2+q'^2}{\Lambda^2} \right) \ ,
\label{eq:ff}
\end{equation}
in the strong interaction kernel, as described in the next section. As explained later, a value of $\Lambda=800$ MeV/$c$ (for all channels) is chosen to fix the position of the quasibound states generated by the solution of the  $T$-matrix equation to observed resonances in some $(S,Q)$ sectors.

\subsection{Strong interaction}

\begin{table}[htbp!]
\resizebox{0.8\columnwidth}{!}{
\begin{tabular}{cccccc}
\hline
 $(S,Q)$ & channel & $a$ & ~~~~~channel~~~~~  & $a_{I_<}$  & $a_{I_>}$ \\
 & (particle)   &  [fm] & (isospin)  &  [fm]  & [fm] \\
\hline\\[-3mm]
$(-1,-1)$ & $D^0 K^-$   & $-0.232$  &  $D{\bar K}$   & $0.399$ & $-0.233$  \\

$(-1,0)$ & $D^0 {\bar K}^0$   & $0.071$    &  & ($I_<=0$) & ($I_>=1$)  \\
             &  ${\boldsymbol{D^+K^-}}$  & $0.083$   &  &  &   \\
           
$(-1,+1)$ & $D^+ {\bar K}^0$   & $-0.233$   &  &    &  \\
            \hline\\[-3mm]           
$(0,-1)$ & $D^0\pi^-$   & $-0.102$    & $D\pi$   &  $0.423 $   & $-0.101$   \\
$(0,0)$ & $D^0\pi^0$  &   $0.056$  &      &  ($I_<=1/2$)  & ($I_>=3/2$)  \\
            &  $\boldsymbol{D^+\pi^-}$   &     $0.253$ & $D\eta$    &  $0.072 + {\rm i}\,0.066$    &    \\
             &  $D^0 \eta $    &  $0.071 + {\rm i}\,0.065$   &  & ($I_<=I_>=1/2$) &  \\
            &  $\boldsymbol{D_s^+K^-}$    & $-0.114+ {\rm i}\,0.693$    &  $D_s {\bar K}$  & $-0.114+ {\rm i}\,0.694$  & \\
$(0,+1)$ & $D^0\pi^+$   & $0.246$    &  & ($I_<=I_>=1/2$)  &   \\
            &  $D^+\pi^0$  & $0.073$    &    & &   \\
             &  $D^+ \eta $    & $0.074 +  {\rm i}\,0.067$    &    &     &  \\
            &  $D_s^+{\bar K}^0$    & $-0.113+ {\rm i}\,0.695$    &    &     &  \\

$(0,+2)$ & $\boldsymbol{D^+\pi^+}$   &  $-0.102$    &    &     &  \\
            \hline \\[-3mm]
            
$(1,0)$ & $\boldsymbol{D_s^+\pi^-}$ &  $0.0033$   &  $D_s\pi$   &    & $0.0032$  \\
              & $D^0K^0$   &  $-0.027 +{\rm i}\,0.084$   &       &   & ($I_>=I_<=1$)  \\
  
 $(1,+1)$ & $D_s^+\pi^0$   & $0.0032$   &   $D K$     &  $-1.28$   &  $-0.027 +{\rm i}\,0.083$ \\
            &  $D^0 K^+$  &  $-0.857+ {\rm i}\,0.020$   &    & ($I_<=0$) &  ($I_>=1$)\\
             &  $D^+ K^0 $    &  $-0.647+ {\rm i}\,0.200$   & $D_s\eta$ & $-0.324 +{\rm i}\,0.132$  &\\
            &  $D_s^+\eta$    &  $-0.324+{\rm i}\,0.132$    &  &  ($I_<=I_>=0$) & \\
            
$(1,+2)$ & $\boldsymbol{D_s^+\pi^+}$ &  $0.0031$   &   &      &  \\
              & $\boldsymbol{D^+K^+}$   &   $-0.026 +{\rm i}\,0.083$   &    &      & \\            
            \hline\\[-3mm]       
$(2,+1)$ & $D_s^+K^0$   & $-0.222$ &  $D_s K$   & $-0.221$ &     \\
$(2,+2)$  & $\boldsymbol{D_s^+K^+}$   & $-0.220$    &   &  ($I_<=I_>=1/2$)  &   \\[1mm]
             \hline
             \hline
  \end{tabular}
  }
 %\centering
 \caption{Two-body channels for definite strangeness $S$ and electric charge $Q$ and corresponding scattering lengths. For each strangeness sector, the scattering lengths in the isospin basis for the lowest $(I_<) $ and highest $(I_>)$ isospin values are also shown.}
 \label{tab:chann}
 \end{table}

The interaction of the $D$-mesons with light particles is described by an effective Lagrangian based on both chiral and heavy-quark symmetries. We use the version at next-to-leading order (NLO) in the chiral expansion, similarly as in Refs.~\cite{Guo:2009ct,Liu:2012zya,Guo:2018tjx,Geng:2010vw,Albaladejo:2016lbb}. In particular we will use our current model described in Ref.~\cite{Montana:2020vjg} for the vacuum zero-temperature case. 

The tree-level scattering amplitude for the interaction of $D$ and $D_s$ mesons with light mesons is given by
\begin{eqnarray} 
 V_{ij}(p_1,p_2,p_3,p_4)&=&\frac{1}{f_\pi^2}\Big[\frac{C_{\rm LO}^{ij}}{4}\left[ (p_1+p_2)^2-(p_2-p_3)^2 \right]-4C_0^{ij}h_0+2C_1^{ij}h_1  \label{eq:potential} \\ 
 &&-2C_{24}^{ij}\Big(2h_2(p_2\cdot p_4)+h_4\big((p_1\cdot p_2)(p_3\cdot p_4)+(p_1\cdot p_4)(p_2\cdot p_3)\big)\Big)\nonumber \\
 &&+2C_{35}^{ij}\Big(h_3(p_2\cdot p_4)+h_5\big((p_1\cdot p_2)(p_3\cdot p_4)+(p_1\cdot p_4)(p_2\cdot p_3)\big)\Big) 
 \Big] \ , \nonumber
\end{eqnarray}
where the $i,j$ indices denote channels with charm $C=1$ and a given value of strangeness $S=\{-1,0,1,2\}$ and total charge $Q=\{-1,0,+1,+2\}$,  $p_1$ and $p_2$ ($p_3$ and $p_4$) are the four-momenta of the mesons in channel $i$ ($j$) and $f_\pi=92.3$ MeV is the pion decay constant. The values of the coefficients $C_{k}^{ij}$, determining the strength of each $k$ term ($k={\rm LO},0,1,24,35$) for the transition $i \to j$ can be found in Appendix A.2 of~\cite{MontanaFaiget:2022cog} for the particle basis employed in the present work. As for the low energy constants (LECs) $h_i$ ($i=0,\dots,5$) of the NLO contributions, we adopt the values of the Fit-2B to LQCD data performed in Ref.~\cite{Guo:2018tjx} and shown in Table~\ref{tab:LECs}.

\vspace{7mm}
\begin{table}[ht!]
\begin{tabular}{|c|c|c|c|c|c|}
\hline
 $h_0$ & $h_1$ & $h_2$ & $h_3$ & $h_4$ & $h_5$ \\
\hline\hline 
0.033 & 0.45 & -0.12 & 1.67 & $-0.0054\cdot 10^{-6}$~MeV$^{-2}$ & $-0.22\cdot 10^{-6}$~MeV$^{-2}$ \\
\hline
\end{tabular}
\centering
\caption{Values of LECs for the $D$-meson--light meson scattering. Taken from Ref.~\cite{Montana:2020vjg}, adapted from Fit-2B in Ref.~\cite{Guo:2018tjx}.}
\label{tab:LECs}
\end{table}

We take the interaction kernel in the center-of-mass (c.m.) frame and retain only its $s$-wave projection, namely
\begin{equation}
 V_{ij}^{\rm s-wave}(p, p^\prime; \sqrt{s}) = \frac{1}{2} \int_{-1}^{1} d\cos \theta_{\boldsymbol{p p'}} \  V_{ij}(p_1,p_2,p_3,p_4) \ ,
 \label{eq:s-wave}
\end{equation} 
where $p\equiv |\boldsymbol{p} |$ ($p^\prime \equiv | \boldsymbol{p}^\prime |$) is the modulus of the momentum of the mesons in the incoming (outgoing) channel in the c.m. frame and $\theta_{\boldsymbol{pp'}}$ is the angle between $\boldsymbol{p}$ and $\boldsymbol{p}^\prime$. We take
$p_{1(2)}=(E_{1(2)},\boldsymbol{p})$ and $p_{3(4)}=(E_{3(4)},\boldsymbol{p}^\prime)$, where the energies are $E=(s+m^2-m^{\prime\,2})/(2\displaystyle\sqrt{s})$ with $m$ being the mass of the meson and $m^\prime$ that of its meson partner in the given channel. This prescription disconnects the momenta of the interacting mesons from the value of the c.m. energy $\sqrt{s}$, hence the former potential can be used in an off-shell calculation.

\subsection{Coulomb interaction}

For channels involving a pair of charged particles one must include the Coulomb force, $V^{\rm C}=\varepsilon \alpha/r$, where $\varepsilon=+1 (-1)$ for identical (opposite) charged particles, $\alpha=1/137$ is the fine structure constant and $r$ the interparticle distance. Our approach consists in adding to the tree-level strong interaction potential of Eq.~(\ref{eq:s-wave}) the effect of the Coulomb interaction in momentum space, following a procedure inspired in that of Refs.~\cite{joachain1975quantum,Holzenkamp:1989tq}. We summarize here the main aspects.

The Coulomb interaction in momentum space is obtained by Fourier transforming the potential in coordinate space,
\begin{equation}
 V^{\rm C}(| \boldsymbol{p}^\prime- \boldsymbol{p} |; \mathcal{ R}_C) =  \int^{\mathcal{ R}_C}_{0} d^3r \ {\rm e}^{ i (\boldsymbol{p}^\prime-\boldsymbol{p}) \cdot \boldsymbol{r} } \ \frac{\varepsilon\alpha}{r} = \frac{4\pi \varepsilon \alpha}{| \boldsymbol{p}^\prime- \boldsymbol{p}  |^2}\left[ 1-\cos( |\boldsymbol{p}^\prime- \boldsymbol{p}  | \mathcal{ R}_C)\right] \ ,
 \label{eq:coul}
\end{equation} 
where only interparticle distances fulfilling $r <  \mathcal{ R}_C$ have been considered in order to regulate its long-range character. The performed truncation renders our momentum space approach numerically tractable, as it avoids the forward scattering singularity of the Coulomb interaction at $|\boldsymbol{p}'-\boldsymbol{p}|=0$. 

From the potential in Eq.~(\ref{eq:coul}) we isolate the $l=0$ component,
\begin{align}
V^{\rm C}_{\rm s-wave}(p,p^\prime; \mathcal{ R}_C) &=  \frac{1}{2} \int_{-1}^{1} d\cos \theta_{\boldsymbol{pp'}} \ V^{\rm C}(|\boldsymbol{p}^\prime- \boldsymbol{p} |; \mathcal{ R}_C) \nonumber \\
&= \frac{2\pi \varepsilon \alpha}{p p^\prime} \left\{ {\rm Ci} \left[ |p^\prime - p| \mathcal{ R}_C \right] - {\rm Ci}\left[ (p^\prime +p) \mathcal{ R}_C \right] + \ln\left(\frac{p^\prime + p}{| p^\prime - p| }\right)\right\}  \ ,
 \label{eq:s-wave_coul}
\end{align} 
where ${\rm Ci}[x]=\int_x^\infty dt \ (\cos t)/t$ is the cosine integral function. We have performed calculations for various values of $\mathcal{ R}_C$ and find our results to stabilize at $\mathcal{ R}_C=60$ fm. In the following, the label $\mathcal{ R}_C$ will be suppressed from the arguments of the Coulomb potential in order to simplify the notation.

Before adding the projected Coulomb interaction to the strong one, two modifications need to be made. These are related to the nonrelativistic character of the Coulomb force, while our approach derives the scattering amplitude $T$ from the
relativistic Bethe-Salpeter equation shown in Eq.~(\ref{eq:Tmat}). First, as the two-body propagator in that equation employs relativistic energies, we implement the following replacement
\be  V^{\rm C}_{\rm s-wave}(p,p^\prime) \longrightarrow
\sqrt{\xi(p;s)} \ V^{\rm C}_{\rm s-wave}(p,p^\prime) \ \sqrt{\xi (p^\prime;s)} \ ,  \ee
where the kinematic factors $\xi$ are given by
\be \xi(p;s)=2 \mu \frac{\sqrt{s}-\omega_1(p)-\omega_2(p)}{ \frac{\lambda(s,m_1,m_2)}{4s}-p^2} \ ,
\ee
with $\mu=m_1 m_2 /(m_1+m_2)$ being the reduced mass and $\lambda(s,m_1,m_2)=[(s-(m_1+m_2)^2][s-(m_1-m_2)^2]$ the K\"all\'en function.
Second, we also need to compensate for the normalization factors of type $1/\displaystyle\sqrt{2\omega}$ tied to the relativistic treatment of the two-body propagator in Eq.~(\ref{eq:Tmat}), which do  not appear in a Lippmann-Schwinger-type formulation appropriate for the nonrelativistic Coulomb interaction. Therefore, the final $s-$wave Coulomb contribution to be added to  the strong interaction kernel (only in diagonal transitions involving a pair of charged mesons) reads,
\be  V^{\rm C,rel}_{\rm s-wave}(p,p^\prime;\sqrt{s}) = 
\sqrt{2\omega_1(p)} \sqrt{2\omega_2(p)} \sqrt{\xi(p;s)} \  V^{\rm C}_{\rm s-wave}(p,p^\prime) \ \sqrt{2\omega_1(p^\prime)} \sqrt{2\omega_2(p^\prime)} \sqrt{\xi(p^\prime;s)}  \ . 
\label{eq:vc_rel}
\ee

Note that the above prescription renders the Coulomb kernel dimensionless, as it should be to match the relativistic structure of Eq.~(\ref{eq:Tmat}).

\section{Results~\label{sec:results}}

\subsection{Generated resonances and scattering lengths}

We first discuss the properties of the scattering amplitudes obtained from Eq.~(\ref{eq:Tmat}) when exclusively the strong interaction kernel of Eq.~(\ref{eq:s-wave}) is employed (no Coulomb). As noted above, the potential includes a form factor [see. Eq~(\ref{eq:ff})] with a cutoff value of $\Lambda=800$~MeV/$c$, which is chosen so as to reproduce the double-pole structure of the $D^*_0(2300)$~\cite{Albaladejo:2016lbb} and to obtain the $D_{s0}^*(2317)$ as a meson-meson bound state. With this prescription, the modulus of the ($S=0$, $I=1/2$) $D\pi$ scattering amplitude presents a broad peak located at $\sqrt{s}=2125$~MeV, while another narrower peak is seen in the modulus of the $D_s \bar{K}$ scattering amplitude at $\sqrt{s}=2462$~MeV. These structures are the reflections in the real energy axis of the two poles in the complex plane representing the $D^*_0(2300)$, found by most unitary meson-meson scattering models and also supported by a recent lattice data analysis~\cite{Asokan:2022usm}. Extending the $T$-matrix equation~(\ref{eq:Tmat}) to complex values of the energy $z \equiv \sqrt{s}$ we are able to numerically find the pole positions. These are represented in Table~\ref{tab:genstates}. Compared to our previous findings of Ref.~\cite{Montana:2020vjg} the results seem compatible to a large extent, given the differences in the approach (on-shell versus off-shell schemes), the different choice of the regulator, and deviations coming from working in the physical versus the isospin basis\footnote{We obtain that the real parts of the poles are separated by approximately 550 MeV. Therefore it might not be meaningful to refer them as a ``two-pole'' state, as argued in Ref.~\cite{Xie:2023cej}.}.

Additionally, the $DK$ amplitude shows to a bound state at 2320~MeV, very close to the nominal value of the $D_{s0}^*(2317)$. Thanks to the heavy-quark spin symmetry, we also find the double pole structure of the $D_1(2430)$ at positions shown in the $J=1, S=0$ sector of Table~\ref{tab:genstates}, which are signaled as maxima in the scattering amplitude at $\sqrt{s}=2267$~MeV and $\sqrt{s}=2606$~MeV, respectively. We also find the bound state $D_{s1}(2460)$ at $\sqrt{s}=2465$~MeV.

%\vspace{7mm}
%\begin{table}[htbp!]
%\begin{tabular}{|c|c|c||c|c|c|}
%\hline 
%\multicolumn{3}{|c||}{$J=0$} & \multicolumn{3}{c|}{$J=1$} \\
%\hline
% Generated state & $(S,I)$ & $\sqrt{s}$ (MeV) & Generated state & $(S,I)$ & $\sqrt{s}$ (MeV) \\
%\hline\hline 
%$D_{0}^{*}(2300)$ (lower pole) & $(0,1/2)$ & 2125 &  $D_{1}(2430)$ (lower pole) & $(0,1/2)$ &  2267 \\
%$D_{0}^{*}(2300)$ (higher pole) & $(0,1/2)$ & 2462 & $D_{1}(2430)$ (higher pole) & $(0,1/2)$ &  2606 \\
%$D_{s0}^{*}(2317)$ & $(1,0)$ & 2320 &  $D_{s1}(2460) $& $(1,0)$ & 2465 \\
%\hline
%\end{tabular}
%\centering
%\caption{Generated states in the pseudoscalar ($J=0$) and vector ($J=1$) sectors.}
%\label{tab:genstates}
%\end{table}

\vspace{7mm}
\begin{table}[htbp!]
\begin{tabular}{|c|c|c||c|c|c|}
\hline 
\multicolumn{3}{|c||}{$J=0$} & \multicolumn{3}{c|}{$J=1$} \\
\hline
 Generated state & $(S,I)$ & $z$ (MeV) & Generated state & $(S,I)$ & $z$ (MeV) \\
\hline\hline 
$D_{0}^{*}(2300)$  (lower pole) & $(0,\frac12)$ & $2092.4+{\rm i}\,129.5$ &  $D_{1}(2430)$ (lower pole) & $(0,\frac12)$ &  $2233.6+{\rm i}\,130.8$ \\
$D_{0}^{*}(2300)$ (higher pole) & $(0,\frac12)$ & $2647.2+{\rm i}\,264.8$  & $D_{1}(2430)$ (higher pole) & $(0,\frac12)$ &  $2719.2+{\rm i}\,330.1$ \\
$D_{s0}^{*}(2317)$ & $(1,0)$ & $2320.2+{\rm i}\,0$ &  $D_{s1}(2460) $& $(1,0)$ & $2464.7 + {\rm i}\,0$ \\
\hline
\end{tabular}
\centering
\caption{Generated states in the pseudoscalar ($J=0$) and vector ($J=1$) sectors as poles of the $T$-matrix of Eq.~(\ref{eq:Tmat}) in the complex energy plane $z=\sqrt{s}$.}
\label{tab:genstates}
\end{table}

Our choice of cutoff produces values of scattering lengths that are in good agreement with different sets in the literature, obtained from chiral unitary models similar to the one adopted here~\cite{Liu:2012zya,Guo:2018tjx,Albaladejo:2016lbb,Guo:2018kno}. The values of our scattering lengths in the different strangeness sectors are shown in Table~\ref{tab:chann}, both in the particle and isospin bases. Note that the scattering lengths in the isospin basis have been obtained employing isospin-averaged masses in each particle multiplet. In the strangeness $S=-1$ sector, we find a moderately large and positive scattering length for the $D{\bar K}$ interaction in $I=0$, representing a sizable attraction, and a negative scattering length of smaller size in $I=1$, representing a moderate repulsion. These isospin-basis values explain the mild attraction found for the $D^+K^-$ interaction, represented by a small positive value of the scattering length. Focusing directly to the pairs of charged mesons, we observe in the strangeness $S=0$ sector a relative important attraction in the $D^+\pi^-$ channel and a milder repulsion in the $D^+_s K^-$ one, as deduced from the size and sign of their corresponding scattering lengths. However, in the later case, this repulsion is apparent, as the negative real part of the scattering length (with a sizable imaginary part) is tied to the influence of the higher pole of the $D^*_0(2300)$, which couples substantially to $D^+_s K^-$. In the $S=1$ sector, the interaction in the channels with a pair of charged mesons is extremely small, slightly attractive for the $D_s^+\pi^-$ and $D_s^+\pi^+$ cases, and somewhat repulsive for the $D^+K^+$ one. In this sector, however, there is a very interesting case, namely that of the channels $D^0 K^+$ and $D^+ K^0$. Their scattering lengths are large and negative, not because the interaction is strongly repulsive, but because it is strongly influenced by the presence of the isospin $I=0$ charm-strange resonance $D_{s0}^*(2317)$ located right below the $D{\bar K}$ threshold.
%Recent calculations~\cite{Albaladejo:2023pzq,Liu:2023uly} obtain a substantial depletion of the $D^0K^+$ correlation function at zero momentum, an effect related to the nearby subthreshold resonance. 
Finally, the $D^+_s K^+$ interaction in the $S=2$ sector is repulsive and of moderate size.

Our latest results of scattering lengths were presented in Ref.~\cite{MontanaFaiget:2022cog}. Compared to those, the present results are compatible but there exist certain deviations in some of the channels. The differences are attributed to several details in the modeling: here we solve the off-shell $T$-matrix equation instead the on-shell version of~\cite{Montana:2020vjg,MontanaFaiget:2022cog}, and here we use a form factor to regularize the integrations instead the hard cutoff used in~\cite{Montana:2020vjg,MontanaFaiget:2022cog}.

\subsection{Coulomb + strong wave functions and matching} 

\begin{figure}[ht!]
\begin{center}
\includegraphics[width=\textwidth]{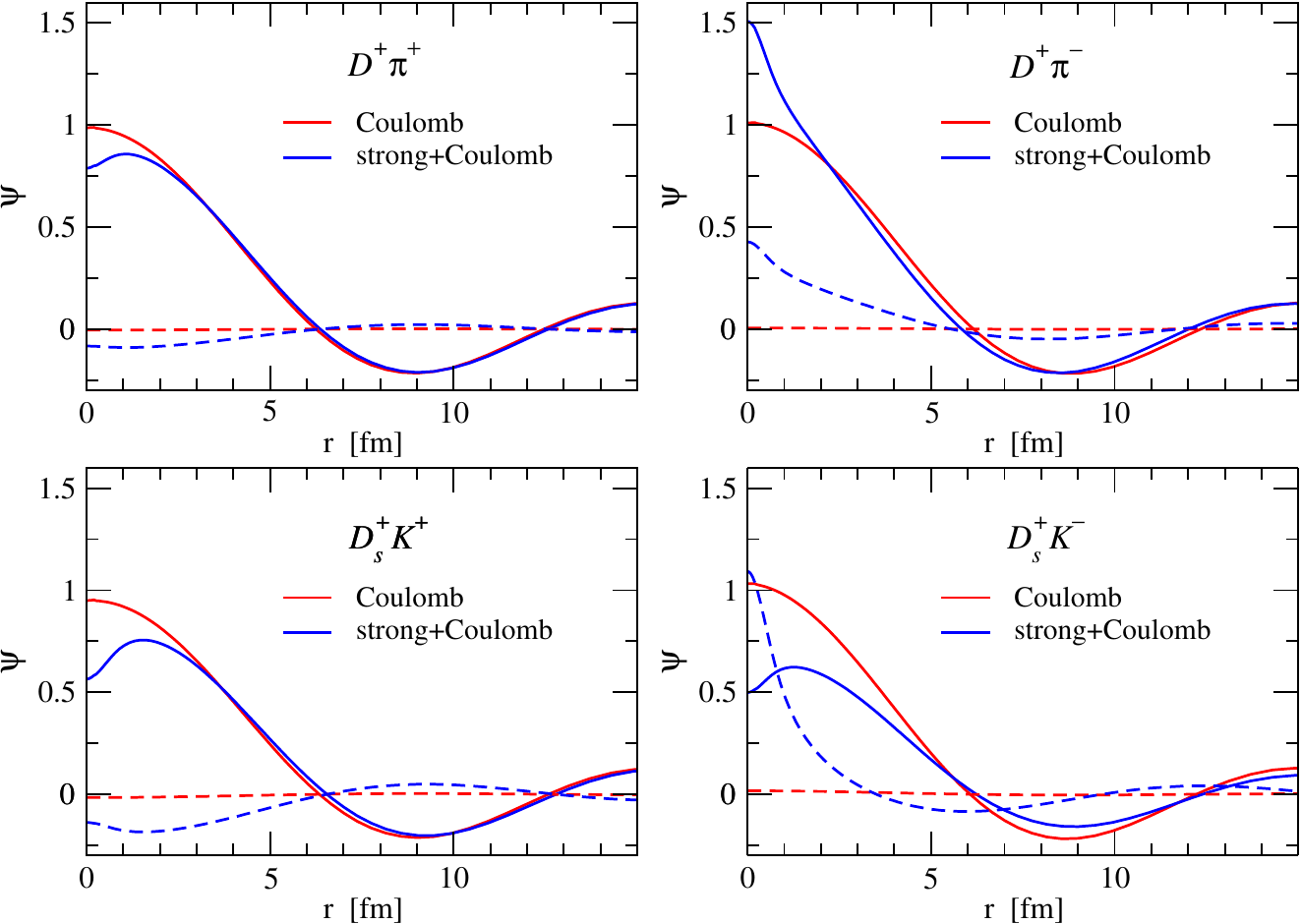}
\end{center}
\caption{$l=0$ wave functions of selected pairs of charged mesons with $q=100$~MeV/$c$. The red curves denote the real (solid) and imaginary (dashed) parts of the Coulomb only wave functions and the blue lines represent the real (solid) and imaginary (dashed) parts of the wave functions that also incorporate the effect of the strong interaction.\label{fig:wf}}
\end{figure}

Taking into account the scattering lengths of Table~\ref{tab:chann}, we consider the cases of pairs of charged mesons that have a more sizable interaction, namely $D^+\pi^+$, $D^+\pi^-$, $D_s^+ K^+$ and $D_s^+ K^-$, and present in Fig.~\ref{fig:wf} their $l=0$ wave functions for a momentum $q=100$~MeV/$c$. The red curves represent the real (solid) and imaginary (dashed) parts of the wave function when only the Coulomb interaction is considered, while the blue lines represent the real (solid) and imaginary (dashed) parts of the wave function when the effect of the strong interaction is also present. Note that the real part of the Coulomb-only wave function at the origin deviates from one, slightly below it for an equal-charge repulsive Coulomb interaction and slightly above it for an opposite-charge attractive one. This deviation is larger for systems with larger reduced masses (bottom panels). The incorporation of the strong force has essentially two visible effects on the complete wave function when compared with the Coulomb-only one, namely a modification around the origin and the addition of a phase-shift at large distances.

We observe that the real part of the wave function at the origin is smaller (larger) than the Coulomb-only counterpart for a repulsive (attractive) strong interaction, as in the case of $D^+\pi^+$ and $D_s^+ K^+$ ($D^+\pi^-$). Note that the relatively large attraction of the strong interaction in the $D^+\pi^-$ channel produces a substantial modification of the wave function at the origin which will be reflected in the corresponding correlation function discussed in the next subsection.
The size of the imaginary part of the full wave function is larger than the Coulomb-only counterpart and follows a similar trend that that seen for the real part. The $D^+_sK^-$ case is special because it is affected by the nearby higher pole of the $D_0^*(2300)$ resonance, which  produces an apparently repulsive amplitude with a sizable imaginary part, as indicated by the scattering length of $-0.11+{\rm i}\,0.69$~fm reported in Table~\ref{tab:chann}. For this reason, the real and imaginary parts of the $D^+_sK^-$ wave function develop substantial differences with respect to the Coulomb-only ones. 

The magnitude of the phase shift acquired by the complete wave function is in accordance to the strength of the strong interaction, being moderate and positive for the repulsive $D^+\pi^+$, $D^+_s K^+$ and $D^+_s K^-$ interactions and larger and negative for the substantially attractive $D^+\pi^-$ case.

\subsection{$D^\pm$-meson correlation functions~\label{sec:D}}

We now present our results on the $D$-meson--light meson correlation functions $C(q)$, focusing on pairs of electrically charged particles, i.e. involving hadrons that can be more easily reconstructed experimentally. Preliminary results for some of these channels have recently been presented by the ALICE collaboration~\cite{Grosa2022,Fabbietti2022,Battistini2023,ALICE:prelim}. We begin with these ones, and then continue showing predictions for the correlation functions involving charged $D_s,D^*,D_s^*$ mesons. The channels containing neutral particles---since they are more difficult to detect experimentally---will be left for a future publication collecting all remaining channels.

We start with the four correlation functions which have already  been considered by the ALICE collaboration~\cite{Grosa2022,Fabbietti2022,Battistini2023,ALICE:prelim}. These are $D^+ \pi^+$, $D^+ \pi^-$, $D^+K^+$, and $D^+K^-$ (plus the respective charge conjugated pairs). Note that the strong and Coulomb interactions are symmetric under $C$-parity and therefore, we only need to consider channels with $D$ mesons (being those with $\bar{D}$ mesons identical under charge conjugation). These pair combinations correspond to channels with strangeness and total charge $(S,Q)=(0,+2),(0,0),(1,+2)$ and $ (-1,0)$, respectively, as can be also seen in Table~\ref{tab:chann}.

\begin{figure}[ht!]
\begin{center}
\includegraphics[width=\textwidth]{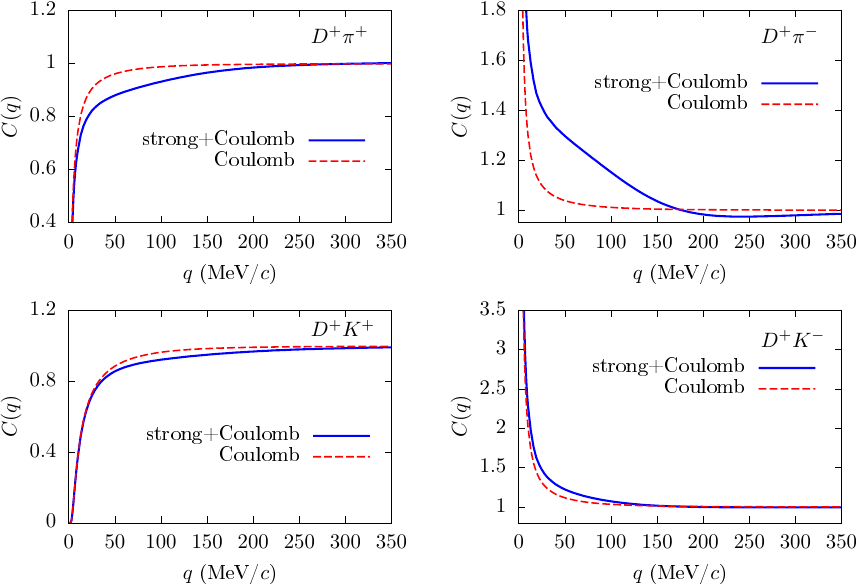}
\end{center}
\caption{Charge $D$-meson--light meson correlation functions as functions of the relative momentum in the center-of-mass reference frame.\label{fig:Dcorr}}
\end{figure}

In Fig.~\ref{fig:Dcorr} we show the correlation functions for the channels $D^+\pi^+,D^+\pi^-,D^+K^+,D^+K^-$ from left to right and from top to bottom, as functions of the relative momentum of the pair. In solid blue line we present our full result including both strong and Coulomb interactions, while the dashed red line contains only the Coulomb effect, for reference.

As expected, the main contribution to the correlation functions is the Coulomb effect, since while having a lesser strength than the strong force, it has a much longer interaction range. For repulsive (attractive) Coulomb interaction the correlation functions are clearly below (above) one. Additionally, we also observe a clear effect of the strong interactions in almost all channels. 

Beginning with the $D^+ \pi^+$ case, we find an additional repulsive interaction. As seen in Table~\ref{tab:chann} this is the only physical channel with $(S,Q)=(0,+2)$ and has a negative scattering length, which is tied to a repulsive leading-order (LO) interaction  ($C_{\textrm{LO}}=1$, see Table~A2 in \cite{MontanaFaiget:2022cog}) for this channel. The strong interaction effects on the $D^+ \pi^-$ correlation function signal a rather pronounced, attractive interaction consistent with a sizable positive value of its scattering length. Note that, while the lowest order coefficient is already of attractive character ($C_{\textrm{LO}}=-1$, see Table~A2 in \cite{MontanaFaiget:2022cog}), there are also sizable $D^+ \pi^-\to D^0 \pi^0$ and $D^+ \pi^- \to D_s^+ K^-$ transitions which, when implemented in a coupled-channel unitarization procedure, enhances the attraction of the  $D^+ \pi^-$ interaction.
This is the case where the double pole of the $D_0^*(2300)$ appears, generated upon unitarization. The lower pole of the $D_0^*(2300)$ is the one that couples more strongly to the $D^+\pi^-$ system and, in the current scheme, leaves a peak at $\sqrt{s} \simeq  2125$ MeV, corresponding to a relative momentum of $q \simeq 200$~MeV/$c$. This pole is likely to produce the shallow minimum below one seen around $q=240$ MeV/$c$, as already noted in Ref.~\cite{Albaladejo:2023pzq} for the $I=1/2$ $D\pi$ channel.
%in the diagonal correlation function (??). The depletion is also seen in Fig.~\ref{fig:Dcorr} around $q=240$ MeV but less pronounced due to the effect of other channels). 
The experimental observation of this minimum would be an evidence of the existence of the lower pole. While the current experimental resolution for this channel would be able to distinguish such a depletion, there exists no current indication of it~\cite{Battistini2023,ALICE:prelim}.

In the bottom left panel of Fig.~\ref{fig:Dcorr} we  consider a channel with net strangeness, the $D^+K^+$ $(S,Q)=(1,+2)$ one, which receives very little influence from the strong interaction. As seen from Table~\ref{tab:chann}, this sector consists of two coupled channels, $D^+K^+$ and $D_s^+ \pi^+$, and the $D^+K^+$ scattering length is complex, with a very small negative real part indicating a tiny repulsion, which is what one can see in the figure. We note that the LO diagonal $D^+K^+ \leftrightarrow D^+K^+$ interaction in this case only starts being nonzero at NLO, meaning a small strength, even after unitarization in coupled channels, which incorporates the effect of the nonzero nondiagonal transition $D^+K^+ \leftrightarrow D_s^+ \pi^+$.

Finally the $D^+ K^-$ correlation in the bottom right panel of Fig~\ref{fig:Dcorr} acquires an additional small increase by the effect of the strong interaction. This $(S,Q)=(-1,0)$ sector has two coupled channels with null diagonal interactions at LO, similarly to the previous case. The non-zero LO non-diagonal transition $D^+K^- \leftrightarrow D^0 \bar{K}^0$ and the NLO terms generate, after unitarization, a small positive scattering length, i.e. the tiny extra attraction seen in Fig.~\ref{fig:Dcorr}. 

%Upon unitarization we observe in Table~\ref{tab:chann} that the final scattering length for this channel is positive, confirming the attractive strong interaction for $D^+\pi^-$. From the same table we see that this sector has four coupled channels, making the repulsion/attraction analysis from the perturbative interaction more cumbersome. In any case, we note that the diagonal (and most relevant) element, $D^+ \pi^- \leftrightarrow D^+ \pi^-$, the leading-order interaction is already attractive (with a negative isospin coefficient). This is also the case for the nondiagonal channels $D^+ \pi^- \leftrightarrow D^0 \pi^0$. 

\begin{figure}[ht!]
\begin{center}
\includegraphics[width=\textwidth]{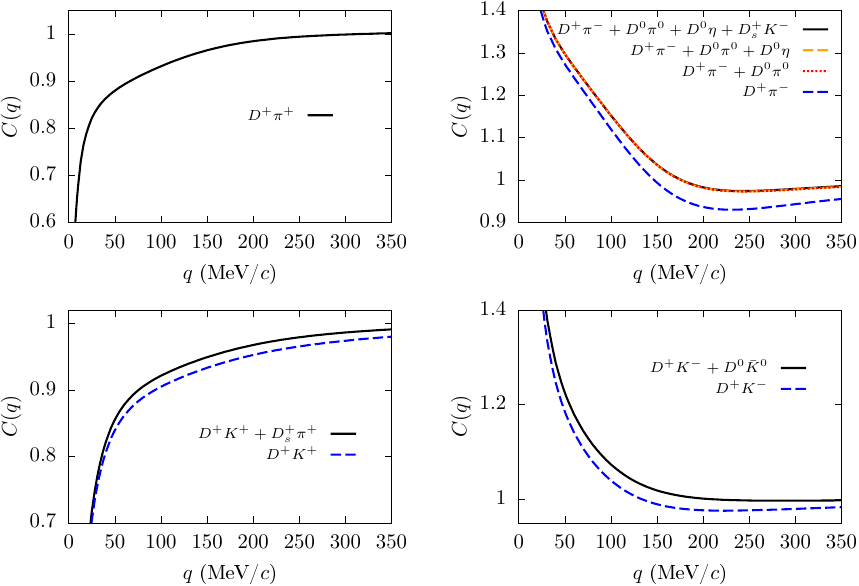}
\end{center}
\caption{$D^+$--light meson correlation functions obtained by sequentially adding the contribution of the different coupled channels of each $(S,Q)$ sector.\label{fig:Dcorr_cc}}
\end{figure}

The effect of coupled channels in the generation of the full the $D^+$--light meson correlation functions is analyzed in Fig.~\ref{fig:Dcorr_cc}. Whereas for the $D^+ \pi^+$ case (top-left panel) there is only one channel, in the top-right panel of Fig.~\ref{fig:Dcorr_cc} the total $D^+ \pi^-$ correlation function is decomposed by adding, sequentially, the individual contributions of the different coupled channels, namely $D^0 \pi^0$, $D^0 \eta$, $D^+_s K^-$, interacting with $D^+ \pi^-$ . We observe that the main contribution to the $D^+ \pi^-$ correlation function comes from the diagonal component of the interaction, and the next most important one is given by the sizable transition $D^+ \pi^- \leftrightarrow D^0 \pi^0$ one. Since these two contributions cause attraction, one could obviously expect a final attractive effect on the correlation function. As for the correlation functions of the $D^+ K^+$ and $D^+ K^-$ pairs (bottom-right and bottom-left panels, respectively) we find again that the most important contribution comes from the diagonal interaction, leading to a slightly larger value for both correlations functions once the nondiagonal contribution is added.

To conclude with the $D-$meson sector we comment on the comparison of our results with the preliminary measurements of the ALICE collaboration presented in Refs.~\cite{Grosa2022,Fabbietti2022,Battistini2023,ALICE:prelim}. There, the experimental results are compared with different theoretical calculations based on the Schr\"odinger equation to match the scattering data of some effective models similar to our own. While our calculation uses the method of the off-shell $T$-matrix calculation in coupled channels, our results show very similar behaviour to other theoretical models in all cases, with a good comparison to experimental data except in the $D^+ \pi^-$ case. The already known discrepancy of the theoretical models with the preliminary experimental data is still present in our case. The depletion we find around $q=240$ MeV/$c$, ascribed to the lower-pole of the $D_0^*(2300)$ state, is not seen in any of the theoretical estimations shown in~\cite{Battistini2023}. The most likely reason is that the (Gaussian) meson-meson potential used in these calculations is fitted to the threshold scattering parameters of the interaction but contains no information about possible subthreshold poles or resonance effects at higher energies.

\subsection{$D_s^\pm$-meson correlation functions~\label{sec:Ds}}

In this section we present our results for the correlation function of the $D_s^+$ meson with the same charged light mesons. We consider the four channels $D_s^+ \pi^+, D_s^+ \pi^-, D_s^+ K^+$ and $D_s^+ K^-$. According to Table~\ref{tab:chann} these channels have quantum numbers $(S,Q)=(1,+2),(1,0),(2,+2)$ and $(0,0)$, respectively.

\begin{figure}[ht!]
\begin{center}
\includegraphics[width=\textwidth]{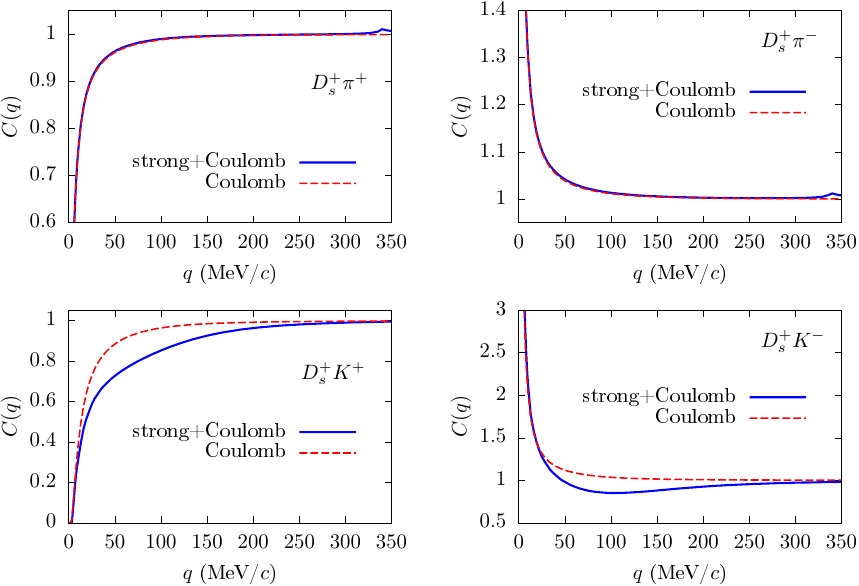}
\end{center}
\caption{$D_s^+$---light meson correlation as functions of the relative momentum in the center-of-mass reference frame.\label{fig:Dscorr}}
\end{figure}

The results of the four correlation functions are plotted in Fig.~\ref{fig:Dscorr} from left to right and top to bottom, in analogy with Fig.~\ref{fig:Dcorr} for the $D^+$ meson case. Again the main contribution to $C(q)$ comes from the Coulomb interaction, being the effect of strong interaction generically smaller than in the case of the $D^+$. 

Starting from the two top panels of Fig.~\ref{fig:Dscorr} we observe a negligible (attractive) effect of the strong interaction in the $D_s^+ \pi^+$ and $D_s^+ \pi^-$ correlation functions, confirmed by the very small positive scattering length in these channels shown in Table~\ref{tab:chann}. These are just two different charged cases of the same $I=1$ $D_s\pi-DK$ coupled-channel interaction, with null LO diagonal transitions.  These channels present a small cusp around $q=340$ MeV/$c$, which is an indication of the opening of the unitary threshold corresponding to the $D^+K^+$ ($D^0K^0$) channel in the $Q=+2$ ($Q=0$) case.

%Similar conclusions can be extracted from the $D_s^+ \pi^-$ correlation function (top-right panel of Fig.~\ref{fig:Dscorr}). This channel has a very similar positive scattering length as $D_s^+ \pi^+$, so that the strong interaction does not appreciably change the attractive Coulomb result. We also expect a cusp happening at the $D^0 K^0$ threshold at $q=340$ MeV which, like the previous case, is very mild to be seen in experimental data.

As opposed to the $D$-meson sector, the correlation functions for the channels involving kaons show a sensible modification due to the strong interaction. In the $D_s^+ K^+$ case (bottom-left panel of Fig.~\ref{fig:Dscorr}) an additional repulsion can be seen in $C(q)$. This is originally caused by the repulsive interaction at LO of the unique channel in this $S=2,Q=+2$ sector, and  confirmed by an appreciably negative scattering length of $a=-0.220$~fm, as seen in Table~\ref{tab:chann}.  Future experimental results looking at this channel might have the needed precision to distinguish between the red and blue results, thus confirming the repulsive nature of the strong interaction in this particular channel.

Finally, for the $D_s^+ K^-$ we observe an additional depletion of the correlation below one with respect to the Coulomb only. This is in contrast to the $D^+ K^-$ case, and a rather interesting effect since the strong interaction in the diagonal channel is attractive at LO. In fact the coupled-channel analysis shows that the real part of the scattering length in the $D_s^+ K^-$ channel is negative. However, this does not indicate repulsion but a strong attraction leading to the presence of a quasibound state that can bring down the correlation function below one producing a shallow minimum, as explained in Ref.~\cite{Liu:2023uly, Albaladejo:2023pzq}.  This is similar to what is observed in the $D^+\pi^-$ correlation function due to the presence of the lower pole of the $D_0^*(2300)$. In the present $D_s^+ K^-$ case the minimum in $C(q)$ is found at $q \simeq 105$ MeV/$c$. Note that the $D_s^+ K^-$ channel--- which is coupled to the $D^+ \pi^-$, where the $D_0^*(2300)$ is found---couples mostly to the higher pole~\cite{Montana:2020vjg}, which appears in the current model at $\sqrt{s} \simeq 2462$ MeV, right below threshold. We conclude that the higher pole of the $D_0^*(2300)$ state is causing the minimum below one of the $D_s^+ K^-$ correlation function. Since this effect is more pronounced that the one in the $D^+ \pi^-$ channel, it would be interesting to test if future precise experimental data involving $D_s$ mesons (maybe from Runs 3 and 4 of the LHC) could resolve the depletion, which would be an evidence of the existence of the higher pole of the $D_0^*(2300)$.

\subsection{$D^{*\pm}$ and $D_s^{*\pm}$ correlation functions~\label{sec:Dstar}}

In this section we briefly comment on the expectation of the correlation functions of heavy vector mesons $D^*, D_s^*$ with light mesons. Heavy-quark spin symmetry allows one to consider the dynamics of vector mesons on the same footing as their pseudoscalar counterparts. Moreover, at LO in the heavy-quark mass expansion the hadron interactions in the pseudoscalar and vector sectors are formally the same, since the interaction with light degrees of freedom are not able to flip the heavy-quark spin, and the $0^-$ and $1^-$ sectors are uncoupled and degenerated. From the point of view of the interaction, the two differences with respect to the pseudoscalar case are the vacuum values of heavy-meson masses, and some of the LECs of the effective Lagrangian at NLO in the chiral expansion~\cite{Montana:2020vjg}. We present the latter in Table~\ref{tab:LECsVector}.

\begin{figure}[ht!]
\begin{center}
\includegraphics[width=\textwidth]{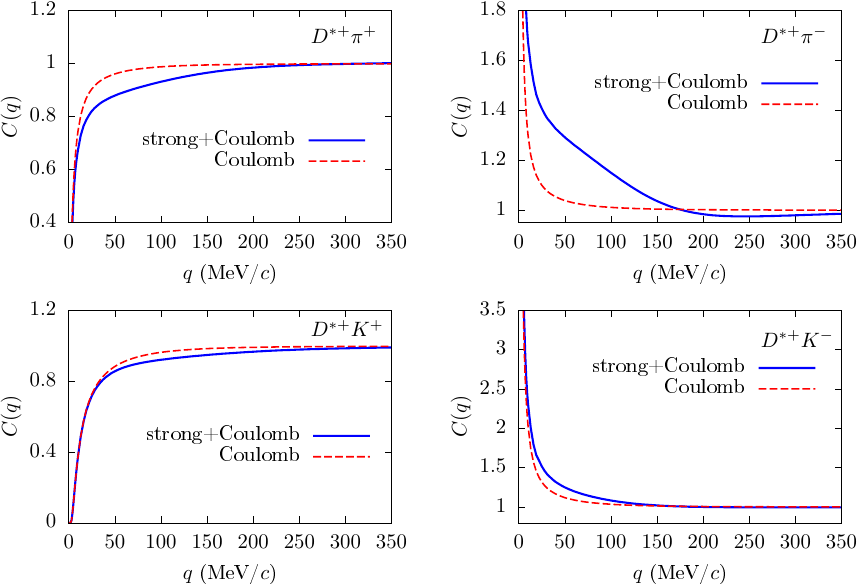}
\end{center}
\caption{Charged $D^*$--light meson correlation functions as functions of the relative momentum in the center-of-mass reference frame.\label{fig:Dstarcorr}}
\end{figure}

\vspace{7mm}
\begin{table}[ht!]
\begin{tabular}{|c|c|c|c|c|c|}
\hline
 $h_0$ & $h_1$ & $h_2$ & $h_3$ & $h_4$ & $h_5$ \\
\hline\hline 
0.033 & 0.45 & -0.12 & 1.67 & $-0.0047\cdot 10^{-6}$~MeV$^{-2}$ & $-0.19\cdot 10^{-6}$~MeV$^{-2}$ \\
\hline
\end{tabular}
\centering
\caption{Values of LECs for the $D^*$--light meson scattering. Taken from Ref.~\cite{Montana:2020vjg}.}
\label{tab:LECsVector}
\end{table}

We first unitarize without modifying the value of the cutoff $\Lambda=800$ MeV/$c$ in the form factors. We automatically generate the spin-partners states to the $D_0^*(2300)$ and the $D_{s0}^*(2317)$, which come out as a prediction of the model: the double pole of the $D_1 (2430)$ and the bound state $D_{s1} (2460)$. The position of the corresponding poles in the complex energy axis  are shown in the $J=1$ sector of Table~\ref{tab:genstates}.

\begin{figure}[ht!]
\begin{center}
\includegraphics[width=\textwidth]{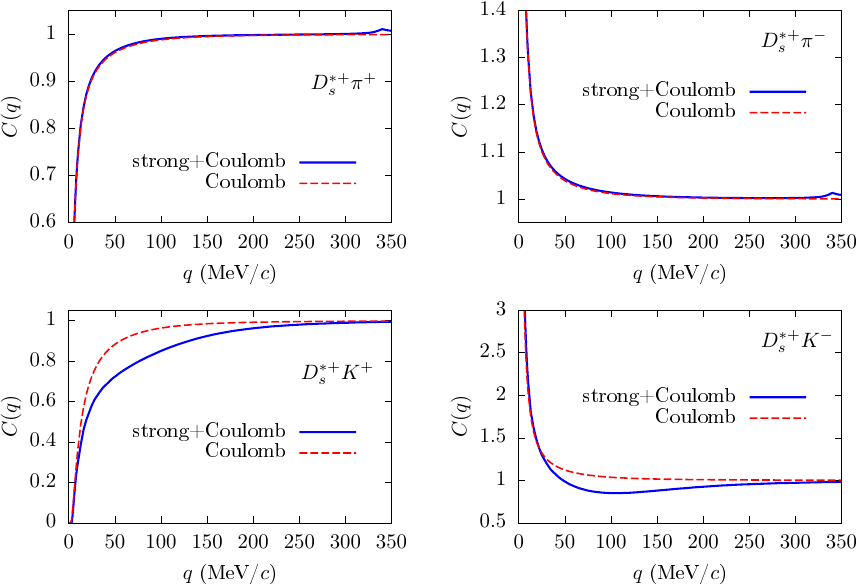}
\end{center}
\caption{Charged $D_s^{*}$--light meson correlation functions as functions of the relative momentum in the center-of-mass reference frame.\label{fig:Dsstarcorr}}
\end{figure}

We present our results for the $D^*$--light meson correlation functions in Fig.~\ref{fig:Dstarcorr}, and those for $D^*_s$--light meson correlation functions in Fig.~\ref{fig:Dsstarcorr}. Despite the differences in the vacuum masses and some of the LECs of the effective Lagrangians, the final results for the correlation functions---when plotted as function of the relative momentum---are essentially indistinguishable from those in the pseudoscalar sectors.

To summarize the most relevant findings in the correlation functions of the $J=1$ sector, we start with the $S=0$ channel, where in $D^{*+}\pi^-$ correlation function of Fig.~\ref{fig:Dstarcorr} we find a shallow depletion around $q=240$ MeV/$c$, which is the imprint of the lower pole of the
$D_1(2430)$ state at an energy $\sqrt{s}=2267$ MeV (or $q \simeq 200$ MeV/$c$). In Fig.~\ref{fig:Dsstarcorr} we find a slightly more pronounced dip in the $D_s^{*+}K^-$ correlation function around $q=100$ MeV/$c$, which is a signature of the higher pole of the $D_1(2430)$ state at $\sqrt{s} =2606$ MeV, right below the channel threshold, therefore behaving as a quasibound state. Hence, these two poles could be observable if experimental data allows for a reasonable precision of these correlation functions involving $D^*$ and $D_s^*$ mesons in the appropriate momentum ranges, perhaps from data taken at Runs 3 and 4 of the LHC. 

Finally we comment that the presence of the $D_{s1} (2460)$ bound state at $\sqrt{s}=2465$ MeV can also leave an imprint in the correlation functions of the $D^{*0} K^+$ or $D^{*+} K^0$. This state is the $J=1$ heavy-quark spin partner of the $D_{s0}^* (2317)$ of the $J=0$ sector for the $D^0 K^+$ and $D^+ K^0$, as was already mentioned in Ref.~\cite{Liu:2023uly}. However, the extraction of the $D^{*0} K^+$ or $D^{*+} K^0$ correlation functions is much more challenging experimentally since the pairs involve neutral mesons.

\section{Conclusions and Outlook~\label{sec:conclu}}

Femtoscopic measurements involving heavy and light mesons have a large potential in terms of capabilities and precision, as has been proven from preliminary ALICE results. We have contributed to this field by calculating femtoscopic correlation functions of charged $D^{(*)}$ and $D_s^{(*)}$ mesons with $\pi^\pm, K^\pm$ mesons. Some of these correlation functions have been already considered by the ALICE collaboration, while some others could be addressed in future analyses. In this work we have argued that several of these correlation functions do contain information about the double-pole structures of the $D_0^*(2300)$ and $D_1(2430)$ states.

Our technique is based on the off-shell $T$-matrix approach, which permits producing scattering amplitudes for the desired channels, given a momentum-dependent potential obtained from an effective field theory and the Coulomb interaction. This has been done in a complete coupled-channel basis and using a Gaussian form factor to tame the ultraviolet divergences of the integrals. The resulting $T$-matrix elements give information about scattering parameters close to thresholds, as well as information about dynamically generated states (resonances and bound states). This resummation method is totally equivalent to solving the corresponding Schr\"odinger equation in coupled channels. With the $T$ matrix, the two-meson wave function can be straightforwardly reconstructed, and then the correlation function via the Koonin-Pratt formula can be obtained.

Our final results for the $D, D_s, D^*$ and $D_s^*$ mesons are shown in Figs.~\ref{fig:Dcorr}, \ref{fig:Dscorr}, \ref{fig:Dstarcorr} and \ref{fig:Dsstarcorr}, respectively. The main conclusions of this work are as follows:

\begin{enumerate}
\item The Coulomb interaction provides the main contribution to the correlation function in all channels, but strong force effects are also clearly visible.

\item Compared to the latest preliminary experimental results by ALICE~\cite{Battistini2023,ALICE:prelim}, the strong interaction is essential to reproduce the data points of the $D^+\pi^+$, $D^+ K^+$ and $D^+K^-$ correlation functions. There remains a puzzle in the $D^+ \pi^-$ correlation function, where our correlation function deviates significantly from the experimental result below $q=350$ MeV/$c$.

\item The lower pole of the $D_0^*(2300)$ state appears as a shallow depletion in the $D^+\pi^-$ correlation function below one around $q = 240$ MeV/$c$. However, preliminary ALICE results---with a rather good precision---do not present evidence for this behavior~\cite{Battistini2023,ALICE:prelim}.

\item The higher pole of the $D_0^*(2300)$ state appears as a depletion in the $D_s^+ K^-$ around $q=100$ MeV/$c$, more pronounced than the one in the $D^+\pi^-$ case. We propose to experimentally analyze the $D_s^+ K^-$ channel to test the presence of this pole.

\item The replacement of $D$ mesons by $D^*$ mesons---following the LO approximation in heavy-meson mass expansion---produces very similar correlation functions as for the $J=0$ case when plotted as functions of the relative momentum. 

\item The lower and higher poles of the $D_1(2430)$ appear as depletions below one around $q=240$ MeV/$c$ and $q=100$ MeV/$c$ in the $D^{*+}\pi^-$ and $D_s^{*+} K^-$ channels, respectively. The analysis of the $D^{*+}\pi^-$ correlation function would be a good way to check whether the mentioned puzzle in the $D^+ \pi^-$ sector still persists in the vector channel. 
\end{enumerate}

In summary, while our $D^+ \pi^-$ correlation function presents significant deviations from preliminary ALICE results, it is unclear which modifications are needed from the theoretical perspective to solve this puzzle. While waiting for definite experimental results in this channel, we propose to look for the $D^{*+} \pi^-$ case to potentially see the effects of the lower pole of the $D_1(2430)$ state. In addition, higher poles in the $J=0$ and $J=1$ sectors couple stronger to $D_s^{+}K^-$ and $D_s^{*+}K^-$, respectively. Then, the study of the correlation functions of these channels would complement the search of the double pole structures of the $D_0^*(2300)$ and $D_1(2430)$, respectively.

Concerning the bound states $D_{s0}^* (2317)$ and $D_{s1} (2460)$, cf. Table~\ref{tab:genstates}, one would require to reconstruct neutral mesons, e.g. look at the $D^0 K^+$ and the $D^{*0} K^+$ channels, respectively. This was already suggested and studied in Refs.~\cite{Albaladejo:2023pzq,Liu:2023uly,Ikeno:2023ojl}. We plan to present our own results with neutral mesons in a separate publication.

Other improvements we would like to address in our model are the use of a more precise form of the source function $S(r)$ (which is straightforward in our code, since it does not need to assume a Gaussian shape), and to revisit the weights $w_i$ of the correlation function. In order to achieve these goals, a deeper knowledge of the experimental details and the particle production mechanism in $p+p$ collisions is mandatory.

\section*{Acknowledgments}

We acknowledge fruitful conversations with L. Fabbietti, F. Grosa, V. Mantovani Sarti, E. Chizzali and D. Battistini.

This research has been supported from the projects CEX2019-000918-M, CEX2020-001058-M (Unidades de Excelencia ``Mar\'{\i}a de Maeztu"), PID2019-110165GB-I00 and PID2020-118758GB-I00, financed by the Spanish MCIN/ AEI/10.13039/501100011033/, as well as by the EU STRONG-2020 project, under the program  H2020-INFRAIA-2018-1 grant agreement no. 824093.  L.T. and J.M.T.-R. acknowledge support from the DFG through project no. 315477589 - TRR 211 (Strong-interaction matter under extreme conditions). L.T. also acknowledges support from the Generalitat Valenciana under contract PROMETEO/2020/023 and from the Generalitat de Catalunya under contract 2021 SGR 171.

\bibliographystyle{ieeetr}
\bibliography{ref.bib}

\begin{thebibliography}{10}

\bibitem{Arndt:2007qn}
R.~A. Arndt, W.~J. Briscoe, I.~I. Strakovsky, and R.~L. Workman, ``{Updated
  analysis of NN elastic scattering to 3-GeV},'' {\em Phys. Rev. C}, vol.~76,
  p.~025209, 2007.

\bibitem{NavarroPerez:2013usk}
R.~Navarro~P\'erez, J.~E. Amaro, and E.~Ruiz~Arriola, ``{Partial Wave Analysis
  of Nucleon-Nucleon Scattering below pion production threshold},'' {\em Phys.
  Rev. C}, vol.~88, p.~024002, 2013.
\newblock [Erratum: Phys.Rev.C 88, 069902 (2013)].

\bibitem{Fabbietti:2020bfg}
L.~Fabbietti, V.~Mantovani~Sarti, and O.~Vazquez~Doce, ``{Study of the Strong
  Interaction Among Hadrons with Correlations at the LHC},'' {\em Ann. Rev.
  Nucl. Part. Sci.}, vol.~71, pp.~377--402, 2021.

\bibitem{HADES:2016dyd}
J.~Adamczewski-Musch {\em et~al.}, ``{The $\bf{\Lambda p}$ interaction studied
  via femtoscopy in p + Nb reactions at $\mathbf{\sqrt{s_{NN}}=3.18}
  ~\mathrm{\bf{GeV}}$},'' {\em Phys. Rev. C}, vol.~94, no.~2, p.~025201, 2016.

\bibitem{Shapoval:2014yha}
V.~M. Shapoval, B.~Erazmus, R.~Lednicky, and Y.~M. Sinyukov, ``{Extracting
  $p\Lambda$ scattering lengths from heavy ion collisions},'' {\em Phys. Rev.
  C}, vol.~92, no.~3, p.~034910, 2015.

\bibitem{STAR:2014dcy}
L.~Adamczyk {\em et~al.}, ``{$\Lambda\Lambda$ Correlation Function in Au+Au
  collisions at $\sqrt{s_{NN}}=$ 200 GeV},'' {\em Phys. Rev. Lett.}, vol.~114,
  no.~2, p.~022301, 2015.

\bibitem{STAR:2015kha}
L.~Adamczyk {\em et~al.}, ``{Measurement of Interaction between Antiprotons},''
  {\em Nature}, vol.~527, pp.~345--348, 2015.

\bibitem{STAR:2018uho}
J.~Adam {\em et~al.}, ``{The Proton-$\Omega$ correlation function in Au+Au
  collisions at $\sqrt{s_{NN}}$=200 GeV},'' {\em Phys. Lett. B}, vol.~790,
  pp.~490--497, 2019.

\bibitem{ALICE:2017jto}
S.~Acharya {\em et~al.}, ``{Measuring K$^0_{\rm S}$K$^{\rm \pm}$ interactions
  using Pb-Pb collisions at ${\sqrt{s_{\rm NN}}=2.76}$ TeV},'' {\em Phys. Lett.
  B}, vol.~774, pp.~64--77, 2017.

\bibitem{ALICE:2017iga}
S.~Acharya {\em et~al.}, ``{Kaon femtoscopy in Pb-Pb collisions at
  $\sqrt{s_{\rm{NN}}}$ = 2.76 TeV},'' {\em Phys. Rev. C}, vol.~96, no.~6,
  p.~064613, 2017.

\bibitem{ALICE:2018nnl}
S.~Acharya {\em et~al.}, ``{Measuring K$^0_{\rm S}$K$^{\rm{\pm}}$ interactions
  using pp collisions at $\sqrt{s}=7$ TeV},'' {\em Phys. Lett. B}, vol.~790,
  pp.~22--34, 2019.

\bibitem{ALICE:2021ovd}
S.~Acharya {\em et~al.}, ``{KS0KS0 and KS0K\ensuremath{\pm} femtoscopy in pp
  collisions at s=5.02 and 13 TeV},'' {\em Phys. Lett. B}, vol.~833, p.~137335,
  2022.

\bibitem{ALICE:2018ysd}
S.~Acharya {\em et~al.}, ``{p-p, p-$\Lambda$ and $\Lambda$-$\Lambda$
  correlations studied via femtoscopy in pp reactions at $\sqrt{s}$ = 7 TeV},''
  {\em Phys. Rev. C}, vol.~99, no.~2, p.~024001, 2019.

\bibitem{ALICE:2020wvi}
S.~Acharya {\em et~al.}, ``{$\Lambda\rm{K}$ femtoscopy in Pb-Pb collisions at
  $\sqrt{s_{\rm{NN}}}$ = 2.76 TeV},'' {\em Phys. Rev. C}, vol.~103, no.~5,
  p.~055201, 2021.

\bibitem{ALICE:2019gcn}
S.~Acharya {\em et~al.}, ``{Scattering studies with low-energy kaon-proton
  femtoscopy in proton-proton collisions at the LHC},'' {\em Phys. Rev. Lett.},
  vol.~124, no.~9, p.~092301, 2020.

\bibitem{ALICE:2021njx}
S.~Acharya {\em et~al.}, ``{Exploring the
  N\ensuremath{\Lambda}\textendash{}N\ensuremath{\Sigma} coupled system with
  high precision correlation techniques at the LHC},'' {\em Phys. Lett. B},
  vol.~833, p.~137272, 2022.

\bibitem{ALICE:2019buq}
S.~Acharya {\em et~al.}, ``{Investigation of the
  p\textendash{}\ensuremath{\Sigma}0 interaction via femtoscopy in pp
  collisions},'' {\em Phys. Lett. B}, vol.~805, p.~135419, 2020.

\bibitem{ALICE:2019eol}
S.~Acharya {\em et~al.}, ``{Study of the $\Lambda$-$\Lambda$ interaction with
  femtoscopy correlations in pp and p-Pb collisions at the LHC},'' {\em Phys.
  Lett. B}, vol.~797, p.~134822, 2019.

\bibitem{ALICE:2019hdt}
S.~Acharya {\em et~al.}, ``{First Observation of an Attractive Interaction
  between a Proton and a Cascade Baryon},'' {\em Phys. Rev. Lett.}, vol.~123,
  no.~11, p.~112002, 2019.

\bibitem{ALICE:2020mfd}
{{ALICE Collaboration}}, ``{Unveiling the strong interaction among hadrons at
  the LHC},'' {\em Nature}, vol.~588, pp.~232--238, 2020.
\newblock [Erratum: Nature 590, E13 (2021)].

\bibitem{ALICE:2020mkb}
S.~Acharya {\em et~al.}, ``{Pion-kaon femtoscopy and the lifetime of the
  hadronic phase in Pb$-$Pb collisions at $\sqrt{s_{\rm{NN}}}$ = 2.76 TeV},''
  {\em Phys. Lett. B}, vol.~813, p.~136030, 2021.

\bibitem{ALICE:2021cpv}
S.~Acharya {\em et~al.}, ``{Experimental Evidence for an Attractive p-$\phi$
  Interaction},'' {\em Phys. Rev. Lett.}, vol.~127, no.~17, p.~172301, 2021.

\bibitem{ALICE:2019igo}
S.~Acharya {\em et~al.}, ``{Measurement of strange
  baryon\textendash{}antibaryon interactions with femtoscopic correlations},''
  {\em Phys. Lett. B}, vol.~802, p.~135223, 2020.

\bibitem{ALICE:2021cyj}
S.~Acharya {\em et~al.}, ``{Investigating the role of strangeness in
  baryon\textendash{}antibaryon annihilation at the LHC},'' {\em Phys. Lett.
  B}, vol.~829, p.~137060, 2022.

\bibitem{ALICE:2022enj}
S.~Acharya {\em et~al.}, ``{First study of the two-body scattering involving
  charm hadrons},'' {\em Phys. Rev. D}, vol.~106, no.~5, p.~052010, 2022.

\bibitem{Grosa2022}
{F. Grosa}, ``{ALICE determines the scattering parameters of D mesons with
  light-flavor hadrons}.'' {Talk on behalf of the ALICE Collaboration, Quark
  Matter 2022, Krak\'ow, Poland}.
\newblock 2022.

\bibitem{Fabbietti2022}
{L. Fabbietti}, ``{D meson scattering parameters with light-flavor hadrons}.''
  {Talk on behalf of the ALICE Collaboration, HF-WINC, Torino, Italy}.
\newblock 2022.

\bibitem{Battistini2023}
{D. Battistini}, ``{Measurement of scattering parameters governing the residual
  strong interaction between charm and light hadrons}.'' {Talk on behalf of the
  ALICE Collaboration, LHCP2023 Conference, Belgrade, Serbia}.
\newblock 2023.

\bibitem{ALICE:prelim}
{ALICE Collaboration}, ``{$D\pi$ and $DK$ femtoscopy in high multiplicity $pp$
  collisions at $\sqrt{s}$=13 TeV }.''
  \url{https://alice-figure.web.cern.ch/node/22039}.
\newblock Accessed: 2023-06-20.

\bibitem{Kamiya:2022thy}
Y.~Kamiya, T.~Hyodo, and A.~Ohnishi, ``{Femtoscopic study on $DD^*$ and
  $D\bar{D}^*$ interactions for $T_{cc}$ and X(3872)},'' {\em Eur. Phys. J. A},
  vol.~58, no.~7, p.~131, 2022.

\bibitem{Vidana:2023olz}
I.~Vidana, A.~Feijoo, M.~Albaladejo, J.~Nieves, and E.~Oset, ``{Femtoscopic
  correlation function for the Tcc(3875)+ state},'' {\em Phys. Lett. B},
  vol.~846, p.~138201, 2023.

\bibitem{Liu:2023wfo}
Z.-W. Liu, J.-X. Lu, M.-Z. Liu, and L.-S. Geng, ``{Distinguishing the spins of
  $P_c(4440)$ and $P_c(4457)$ with femtoscopic correlation functions}.''
  {arXiv: 2305.19048 [hep-ph]}.

\bibitem{Albaladejo:2023pzq}
M.~Albaladejo, J.~Nieves, and E.~Ruiz-Arriola, ``{Femtoscopic signatures of the
  lightest S-wave scalar open-charm mesons},'' {\em Phys. Rev. D}, vol.~108,
  no.~1, p.~014020, 2023.

\bibitem{Liu:2023uly}
Z.-W. Liu, J.-X. Lu, and L.-S. Geng, ``{Study of the DK interaction with
  femtoscopic correlation functions},'' {\em Phys. Rev. D}, vol.~107, no.~7,
  p.~074019, 2023.

\bibitem{Ikeno:2023ojl}
N.~Ikeno, G.~Toledo, and E.~Oset, ``{Model independent analysis of femtoscopic
  correlation functions: An application to the $D_{s0}^*(2317)$}.'' {arXiv:
  2305.16431 [hep-ph]}.

\bibitem{Koonin:1977fh}
S.~E. Koonin, ``{Proton Pictures of High-Energy Nuclear Collisions},'' {\em
  Phys. Lett. B}, vol.~70, pp.~43--47, 1977.

\bibitem{Pratt:1990zq}
S.~Pratt, T.~Csorgo, and J.~Zimanyi, ``{Detailed predictions for two pion
  correlations in ultrarelativistic heavy ion collisions},'' {\em Phys. Rev.
  C}, vol.~42, pp.~2646--2652, 1990.

\bibitem{joachain1975quantum}
C.~Joachain, {\em Quantum Collision Theory}.
\newblock North-Holland Publishing Company, 1975.

\bibitem{Guo:2009ct}
F.-K. Guo, C.~Hanhart, and U.-G. Meissner, ``{Interactions between heavy mesons
  and Goldstone bosons from chiral dynamics},'' {\em Eur. Phys. J.}, vol.~A40,
  pp.~171--179, 2009.

\bibitem{Liu:2012zya}
L.~Liu, K.~Orginos, F.-K. Guo, C.~Hanhart, and U.-G. Meissner, ``{Interactions
  of charmed mesons with light pseudoscalar mesons from lattice QCD and
  implications on the nature of the $D_{s0}^*(2317)$},'' {\em Phys. Rev.},
  vol.~D87, no.~1, p.~014508, 2013.

\bibitem{Guo:2018tjx}
Z.-H. Guo, L.~Liu, U.-G. Mei{\ss}ner, J.~A. Oller, and A.~Rusetsky, ``{Towards
  a precise determination of the scattering amplitudes of the charmed and
  light-flavor pseudoscalar mesons},'' {\em Eur. Phys. J.}, vol.~C79, no.~1,
  p.~13, 2019.

\bibitem{Geng:2010vw}
L.~S. Geng, N.~Kaiser, J.~Martin-Camalich, and W.~Weise, ``{Low-energy
  interactions of Nambu-Goldstone bosons with $D$ mesons in covariant chiral
  perturbation theory},'' {\em Phys. Rev.}, vol.~D82, p.~054022, 2010.

\bibitem{Albaladejo:2016lbb}
M.~Albaladejo, P.~Fernandez-Soler, F.-K. Guo, and J.~Nieves, ``{Two-pole
  structure of the $D^\ast_0(2400)$},'' {\em Phys. Lett.}, vol.~B767,
  pp.~465--469, 2017.

\bibitem{Montana:2020vjg}
G.~Monta\~na, A.~Ramos, L.~Tolos, and J.~M. Torres-Rincon, ``{Pseudoscalar and
  vector open-charm mesons at finite temperature},'' {\em Phys. Rev. D},
  vol.~102, p.~096020, 7 2020.

\bibitem{MontanaFaiget:2022cog}
{G. Monta\~na Faiget}, ``{Effective-theory description of heavy-flavored
  hadrons and their properties in a hot medium}.'' {Ph.D. thesis. Universitat
  de Barcelona (2022). arXiv: 2207.10752}.

\bibitem{Holzenkamp:1989tq}
B.~Holzenkamp, K.~Holinde, and J.~Speth, ``{A Meson Exchange Model for the
  Hyperon Nucleon Interaction},'' {\em Nucl. Phys. A}, vol.~500, pp.~485--528,
  1989.

\bibitem{Asokan:2022usm}
A.~Asokan, M.-N. Tang, F.-K. Guo, C.~Hanhart, Y.~Kamiya, and U.-G. Mei\ss{}ner,
  ``{Can the two-pole structure of the $D_{0}^{*}(2300)$ be understood from
  recent lattice data?},'' {\em Eur. Phys. J. C}, vol.~83, no.~9, p.~850, 2023.

\bibitem{Xie:2023cej}
J.-M. Xie, J.-X. Lu, L.-S. Geng, and B.-S. Zou, ``{Two-pole structures as a
  universal phenomenon dictated by coupled-channel chiral dynamics}.'' {arXiv:
  2307.11631 [hep-ph]}.

\bibitem{Guo:2018kno}
X.-Y. Guo, Y.~Heo, and M.~F.~M. Lutz, ``{On chiral extrapolations of charmed
  meson masses and coupled-channel reaction dynamics},'' {\em Phys. Rev. D},
  vol.~98, no.~1, p.~014510, 2018.

\end{thebibliography}

\end{document}